\documentclass[prd,preprint]{revtex4}
\usepackage{graphicx}

\begin{document}

\title{Factorization and Resummation in Soft-Collinear Effective Theory\footnote{Talk given by Chong Sheng Li
at the Workshop on the Frontiers of Theoretical Physics and
Cross-Disciplinary, NFSC, P.R.China, Beijing, March 2005,
published in the Proceedings, 124-136}}
\author{Yang Gao}
\author{Chong Sheng Li}
\author{Jian Jun Liu}
\affiliation{Department of Physics, Peking University, Beijing
100871, China}
\date{\today}

\begin{abstract}
We review soft-collinear effective theory (SCET), and apply it to
discuss quark electromagnetic form factor, then present the
resumed transverse momentum distribution of Higgs-boson production
via gluon fusion under this framework, where we derive a
relatively full differential formula in transverse momentum $Q_T$
space like one which have been obtained by
Dokshitzer-D'Yanov-Troyan (DDT) in perturbative Quantum
Chromodynamics (pQCD). Furthermore, our above result can be
generalized to even higher order. Comparing our formula with the
integral formula of Collins-Soper-Sterman (CSS) in impact
parameter $b$ space, we establish the relationship between the
anomalous dimension of operator together with matching
coefficients in SCET and the well-known coefficients A, B and C in
pQCD, which also provides a relative natural and convenient method
to treat the similar questions as ones of CSS, such as the
matching in nonperturbative region. Finally, the joint resummation
method in SCET is briefly discussed.
\end{abstract}

\maketitle

\section{Introduction}

Quantum Chromodynamics (QCD) is a highly nonlinear gauge field
theory. Comparing with Quantum Electrodynamics (QED), QCD has very
complex mass singularity due to the self-interactions of gluons
and the failure of perturbation at low energy region. However, QCD
is a field theory with asymptotic freedom, which means
perturbative approach is trustful within high energy region. Thus
factorization, which separates long distance or nonperturbative
and short distance or perturbative contributions of a process,
i.e. subtracting infrared (IR) divergences from observable,
particularly collinear divergences, is necessary to apply QCD to
high energy hard processes. Moreover, when in a process there are
two separating scales, for example, the distribution in the edge
of phase space, double logarithmic (DL) terms would appear that
may spoil perturbative expansion. Thus, we need resummation
technique to control such behavior.

There exist two methods to satisfy above requirements. The first
one is perturbative QCD (pQCD) approach based on the analysis of
Feynman diagrams: Collins-Soper-Sterman (CSS) formulism
\cite{kt,ktt}; the other one is effective field theory (EFT)
approach based on Lagrangian and operators \cite{lb,hsf}. The
spirit of factorization in pQCD approach can be briefly summarized
as follows: first find leading order (LO) IR divergence of Feynmam
diagrams through Landau equation; then separate these IR
divergences into soft and collinear parts by utilizing gauge
invariance and unitarian relations, thus while soft divergences
are cancelled, the left collinear divergences only remain in
initial states; at last absorb them into jet functions. As for
resummation, because conventional renormalization group equation
(RGE) can not resum the terms such as DL between two scales, pQCD
utilizes a new type of equations which is derived by the
differentiation of jet-function from the factorized cross section
with respect to the axial parameter in axial gauge, in order to
disentangle soft and collinear contributions which are origins of
DL terms, and so DL terms can be resummed by solving this
equation.

However, the proof of factorization in pQCD is tedious, and
resummation in pQCD is inconvenient as well. Recently,
soft-collinear effective theory (SCET) has made great
simplifications on the proof of factorization in B meson decays
\cite{bsr,lb} and high energy hard scattering processes
\cite{ee,hsf}, including resummation of large logs in certain
regions of phase space, for example, $e^{+}e^{-}$ annihilation
into two jets of thrust $T\rightarrow1$ \cite{ee,eet}, the deep
inelastic scattering (DIS) in the threshold region $x\rightarrow1$
\cite{dis} and Drell-Yan (DY) process in the case of
$z\rightarrow1$ \cite{dy}. The reason is that SCET can be viewed
as an operator realization of the pQCD analysis when the modes
participating the interactions of interest are soft and collinear,
just like chiral dynamics vs QCD at low energy region. EFT
provides a simple and systematic method for factorization of hard,
collinear and usoft or soft degrees of freedom at operator level,
especially usoft modes can be decoupled from collinear modes by
making a field redefinition, and large logarithms such as
$\log(\mathrm{Q}^2/\Lambda^{2})$, where $\mathrm{Q},\Lambda$ are
two typical scales that character the processes, can be resummed
naturally through RGE running.

In the following, after a brief introduction to SCET, we will
discuss quark form factor and resummed $Q_T$ distributions of
Higgs-boson production via gluon fusion in small $Q_T$ region
within the framework of SCET \cite{ktn}, which was studied using
CSS formulism in pQCD before.

\section{Review of SCET}

SCET is appropriate for the kinematic regions of collinear and
usoft(soft) modes with momenta scaling:
$p_{c}=(p^{+},p^{-},p_{\perp})=(n\cdot p,\bar{n}\cdot
p,p_{\perp})\sim Q(\lambda^{2},1,\lambda)$ and $p_{us}\sim
Q(\lambda^{2},\lambda^{2},\lambda^{2})$ or $p_{s}\sim
Q(\lambda,\lambda,\lambda)$, where the light-like vectors
$n,\bar{n}$ satisfy $n\cdot\bar{n}=2$ and the perpendicular
components of any four vector $V$ are defined by
$V_{\perp}^{\mu}=V^{\mu}-(n\cdot
V){\bar{n}^{\mu}}/{2}-(\bar{n}\cdot V){n^{\mu}}/{2}$ and in the
rest of the report the common scale $Q$ is often omitted. As any
other EFT, SCET should reproduce the infrared behavior of the full
theory, which is ensured by using the method of regions for
Feynman integrals, i.e., by expanding the integrand in different
momentum regions which contribute to the integrals \cite{sr}. This
fact provides us a direct matching calculation to determine the
Wilson coefficients and anomalous dimensions of the operators in
SCET. Because the matching procedure is independent of
regularization, we can calculate on-shell scaleless matrix
elements and use dimensional regularization to regulate UV and IR
divergences, and in this case the loop integrals in SCET are zero,
and results in
IR$_{\mathrm{\mathrm{QCD}}}=\mathrm{IR}_{\mathrm{SCET}}=-\mathrm{UV}_{\mathrm{SCET}}$.

In constructing SCET, one shouold first identify all possible
modes in initial and final states, then with them construct all
possible propagator and vertex. In actual application, there
always involves two kinds of SCET \cite{rpi}:

$\bullet$ SCET$_{II}$: soft and collinear modes for exclusive or
semi-inclusive processes, such as $B \to \pi\nu e, \pi\pi, D\pi$
with momentum scaling as
$$B,D\sim\mathcal{O}(\lambda,\lambda,\lambda),\qquad\
\pi\sim\mathcal{O}(\lambda^2,1,\lambda),$$ where
$\lambda=\frac{\Lambda}{Q}$.

$\bullet$ SCET$_{I}$: usoft and collinear modes for inclusive
processes, say $B \to X_s^*\gamma$ at the end point region and
$e^-p\to e^-\mathrm{X}$ at the threshold region with momentum
scaling as
$$B\sim\mathcal{O}(\lambda^{2},\lambda^{2},\lambda^{2}),\qquad\
X_s^*, \mathrm{X}\sim\mathcal{O}(\lambda^2,1,\lambda),$$ where
$\lambda=\sqrt{\frac{\Lambda}{Q}}$. We notice that $SCET_{II}$ can
be also viewed as an EFT of $SCET_{I}$, i.e., $SCET_{I}\rightarrow
SCET_{II}$. Hence we first discuss $SCET_I$, at LO of $\lambda$,
the collinear parts of lagrangian in $SCET_{I}$ are
$\mathcal{L}_{c}=\mathcal{L}_{cq}+\mathcal{L}_{cg}$, where
$$\mathcal{L}_{cg}=\frac{1}{2g^2}tr\{[i\mathcal{D}^{\mu}+gA_{n,q}^{\mu},i\mathcal{D}^{\nu}
+gA_{n,q'}^{\nu}]\}^2$$
$$+2tr\{\bar{c}_{n,p'}[i\mathcal{D}_{\mu},[i\mathcal{D}^{\mu}+
A^{\mu}_{n,q},c_{n,p}]]\}+\frac{1}{\alpha}tr
\{[i\mathcal{D}_{\mu},A_{n,q}^{\mu}]\}^2 \sim
\mathcal{O}(\lambda^{4}),$$ here the last is the gauge fixing term
with parameter $\alpha$ and $c_{n,p}$ denotes collinear ghost
field, and
$$\mathcal{L}_{cq}=\bar{\xi}_{n,p'}[in\cdot
D+i\not{D}_{\perp}^c\frac{1}{i\bar{n}\cdot D^c}i\not{D}_{\perp}^c]
\frac{\not{\bar{n}}}{2}\xi_{n,p}\sim \mathcal{O}(\lambda^{4}),$$
with
$$in\cdot D=in\cdot D_{us}+in\cdot A_n, \qquad\
iD_{us}=i\partial+gA_{us},$$
$$i\bar{n}\cdot D^c=\bar\mathcal{{P}}+g\bar{n}\cdot A_n,\qquad\
iD_{\perp}^c=\mathcal{P}_{\bot}+gA_n^\perp.$$ The power counting
in SCET is straightforward once the scaling of fields is obtained
from the scaling of their corresponding propagators \cite{sr}. The
collinear fermion field is a two component spinor in the $n$
direction, after its small components integrated out by solving
the equation of motion, and can be expanded as
\begin{equation}
\xi_{n}(x)=\sum_{\tilde{p}}e^{-i\tilde{p}\cdot x}\xi_{n,p}(x),
\end{equation}
where $p=\tilde{p}+\mathcal{O}(\lambda^{2})$ and
$n\cdot\tilde{p}=0 $, $\tilde{p}$ is called label momentum. The
order $\lambda^2$ components are resided in the space-time
dependence of the fields $\xi_{n,p}(x)$. The label operators
$\bar\mathcal{{P}}$ and $\mathcal{P}_{\bot}$ pick out
$\mathcal{O}(1)$ and $\mathcal{O}(\lambda)$ label momentum of
collinear fields respectively, which is introduced to facilitate
power counting in SCET and conservation of momenta \cite{hsf}. The
label operator
$\mathcal{P}^{\mu}=\bar{\mathcal{P}}\frac{n^\mu}{2}+\mathcal{P}_{\bot}^\mu$
is defined by
$$\mathcal{P}^{\mu}(\phi_{q}^\dagger \cdots \phi_{p} \cdots)
=(\tilde{p}^{\mu}+\cdots-\tilde{q}^{\mu}-
\cdots)\times(\phi_q^\dagger \cdots \phi_p \cdots)$$ for collinear
fields and
$$i\mathcal{D}^{\mu}=\bar\mathcal{P}\frac{n^\mu}{2}+
\mathcal{P}_{\bot}^\mu+(in\cdot\partial+gn\cdot A_{us})
\frac{\bar{n}^\mu}{2}.$$ The lagrangian of usoft modes is the same
as that of QCD.

There are three types of symmetries in SCET. The first one is of
spin structures, as the collinear quark fields are two-component
spinors, the independent Dirac matrixes are
\{$\not{\bar{n}},\not{\bar{n}}\gamma_5,\not{\bar{n}}\gamma_\perp^\mu$\}
or \{$1,\gamma_5,\gamma_{\perp}^\mu$\}. The second is
re-parameters(RP) transformations invariance under the
infinitesimal transformations:

$\bullet$ $n\to n+\triangle_{\perp},\qquad\bar{n}\to \bar{n}$;

$\bullet$ $n\to n,\qquad\bar{n}\to \bar{n}+\epsilon_{\perp}$;

$\bullet$ $n\to e^{\alpha}n,\qquad\bar{n}\to e^{-\alpha}\bar{n}$.\\
The requirement of
$\triangle_{\perp}\sim\lambda,\epsilon_{\perp}\sim\lambda^0,\alpha\sim\lambda^0$
is to keep scaling behavior of the elements in $SCET_{I}$ under
these transformations. The third symmetry is invariance of gauge
transformations:

$\bullet$ Collinear transformation $U_c$, $\partial^\mu U_{c}\sim
\mathcal{O}(\lambda^{2},1,\lambda)$ and $\partial^\mu U_{c,q}\sim
\mathcal{O}(\lambda^{2},\lambda^{2},\lambda^{2})$
$$\xi_{n}\to U_c\xi_{n},\qquad\ W_{n}\to U_cW_{n},$$
$$A_{n}\to U_cA_{n}U_c^\dagger+\frac{i}{g}U_c[i\mathcal{D},U_c^\dagger],$$
while usoft operators do not transform.

$\bullet$ Usoft transformation $U_{us}$, $\partial^\mu U_{us}\sim
\mathcal{O}(\lambda^{2},\lambda^{2},\lambda^{2})$
$$\xi_{n}\to U_{us}\xi_{n},\qquad\ W_{n}\to U_{us}W_{n}U_{us}^\dagger,$$
$$A_{n}\to U_{us}A_{n}U_{us}^\dagger,$$ while
usoft operators transform as in QCD. Where
\begin{equation}
W_{n}(x)=[\sum_{\mathrm{perms}}\exp(\frac{-g}
{\bar\mathcal{{P}}}\bar{n}\cdot A_{n,q})]
\end{equation}
denotes a Wilson line of collinear gluons along the path in the
$\bar{n}$ direction, which is required to ensure gauge invariance
of current operator in SCET; $Y_{n}(x)=\mathrm{P}\exp(ig\int
dsn\cdot A_{us}(ns+x))$ is usoft Wilson line of usoft gluons in
$n$ direction from $s=0$ to $s=\infty$ for final state particles
and $\mathrm{P}$ means path-ordered product, while for initial
state particles, $Y_{n}$ is from $s=-\infty$ to $s=0$.

The above properties result in no extra-renormalization theorem of
$\mathcal{L}$ and connecting Wilson coefficients of different
operators \cite{sr}. Until now the lagrangian of SCET is
completely constructed, next we also need to determine various
effective operators in SCET through matching to count into
contributions from high energy region.

For example, we match $heavy \to light$ current $b\to s\gamma$ at
end point region \cite{bsr}, in following we take the convention
$\phi_{n,p}(x)=\phi_{n}(x)$ for collinear fields and use
$\overline{\mathrm{MS}}$ scheme, i.e., $\mu^2\rightarrow
\mu^2e^{\gamma_E}/4\pi$,
$$J^{QCD}=\bar{s}\Gamma b \rightarrow J^{SCET_{I}}=[\bar{\xi}_nW_{n}]\Gamma h_v$$
at the scale $\mu=p^-\sim m_b$,
$$J^{QCD}=(\bar{\xi}_{n}W_{n})_{p^-}\Gamma h_vC(p^{-},\mu)=
[\bar{\xi}_{n}W_{n}]C(\bar{\mathcal{P}}^\dagger,\mu)\Gamma h_v,$$
where $C(m_b)=1+\mathcal{O}(\alpha_s)$, and
$$Z=1+\frac{\alpha_sC_F}{4\pi}[\frac{1}{\epsilon^2}+\frac{1}{\epsilon}
\ln(\frac{\mu^2}{m_b^2})+\frac{5}{2\epsilon}].$$ Here
$C_F=\frac{N_c^2-1}{2N_c}$ for $SU(N_c)$ QCD with $N_c=3$.

After above matching procedure, we should running the operator in
$SCET_{I}$ from scale $\mathcal{O}(Q)$ to
$\mathcal{O}(\sqrt{Q\Lambda})$ by RGE, so as to control the large
logarithmic behavior. RGE at next-to-leading-logarithmic-order
(NLLO) thus is given by
$$\mu\frac{d}{d\mu}C(\mu)=-\frac{\alpha_s(\mu)C_F}{2\pi}[2\ln(\frac{\mu}{m_b})+
\frac{5}{2}]C(\mu).$$ As to the meaning of this RGE, we can set
$\alpha_s=\alpha,C_F=1\sim QED$ at LLO
$$C(\mu)=\exp[-\frac{\alpha}{2\pi}\ln^2(\frac{\mu}{\bar{p}})].$$
So it is indicated that Sudakov's DL can be resummed through
conventional RGE, which is different from RGE in full theory. Thus
EFT provides a natural method for resummation.

Next, we turn to the factorization in terms of SCET. The
factorization in SCET is represented as decoupling of collinear
and usoft modes in $\mathcal{L}$ and relevant operators, which can
be accomplished by the following decoupling transformations for
initial state particles:
$$\xi_{n,p}=Y\xi_{n,p}^{(0)},\qquad\
A_{n,p}=YA_{n,p}^{(0)}Y^\dagger,\qquad\
W_{n}=YW_{n}^{(0)}Y^\dagger.$$ Therefore, $\mathcal{L}$ in terms
of $\xi_{n,p}^{(0)}$ and $A_{n,p}^{(0)}$ can be written as
$$\mathcal{L}=\bar{\xi}_{n,p'}\frac{\not{\bar{n}}}{2}
[in\cdot\mathcal{D}+gn\cdot A_{n,q}+\cdots]\xi_{n,p}$$
$$=\bar{\xi}_{n,p'}^{(0)}\frac{\not{\bar{n}}}{2}[in\cdot\partial+gn\cdot
A_{n,q}^{(0)}+\cdots]\xi_{n,p}^{(0)}$$ or
$\mathcal{L}(\xi_{n,p},A_{n,q},n\cdot
A_{us})=\mathcal{L}(\xi_{n,p}^{(0)},A_{n,q}^{(0)},0)$, which is
already a decoupled form. And whether an operator has the
decoupled form depends on the case.

If we want to go to $SCET_{II}$, the following steps should be
performed:

(i) Matching QCD onto SCET$_{I}$ at a scale $\mu^{2}\sim Q^{2}$
with $p_{c}^{2}\sim Q^{2}\lambda^{2}$.

(ii) Decoupling the usoft -collinear interactions with the field
redefinitions, $\xi_{n}=Y_{n}^{\dagger}\xi_{n}^{(0)}$ and
$A_{n}=Y_{n}^{\dag}A_{n}^{(0)}Y_{n}$ for final state particles,
while for initial states particles the daggers are reversed.

(iii) Matching SCET$_{I}$ onto SCET$_{II}$ at a scale $\mu^{2}\sim
Q^{2}\eta^{2}$ with $p_{c}^{2}\sim Q^{2}\eta^{2}$, where $\eta\sim
\lambda^{2}$.

Considering heavy to light current, for example, it leads to
$$J^{QCD}=\bar{\psi}\Gamma h\to J^{SCET_{I}}=\bar{\xi_n}W_{n}\Gamma
h_v;$$ $$J^{SCET_{I}}=\bar{\xi_n}W_{n}\Gamma
h_v=\xi_n^{(0)}W_{n}^{(0)}\Gamma Y^\dagger h_v;$$and
$$J^{SCET_{I}}=\bar{\xi}_{n}^{(0)}W_{n}^{(0)}\Gamma Y^\dagger h_v
\to J^{SCET_{II}}=\bar{\xi}_nW_{n}\Gamma S^\dagger h_v.$$

We thus can obtain general factorized convolutional structure of
any quantity $\sigma $ from matching:
$$\sigma \sim H\otimes J \otimes S.$$ The function $H$ encodes the short distance
physics, the jet function $J$ describes the propagation of
energetic particles in collimated jets, and the soft function $S$
contains relative long-distance physics.

\section{quark electromagnetic form factor in QCD \cite{hsf,ktn}}

As a demonstration of the statements in last section, we consider
Sudakov effect of quark electromagnetic form factor in QCD
\cite{onfsf,ofsf}, i.e., double logarithmic asymptotic of current
$j^{\mu}=\bar{\psi}\gamma^{\mu}\psi$ in the following kinematics:

\ \ \hspace{1.2cm}(a) on-shell case
$$Q^{2}=-(p-k)^{2}>>\Lambda^{2}>>m^{2}, \qquad\
p^{2}=k^{2}=m^{2}.$$

\ \ \hspace{1.2cm}(b) off-shell case
$$Q^{2}=-(p-k)^{2}>>-p^{2}=-k^{2}=M^{2}>>m^{2}.$$
where $p,k$ are momenta of an initial and a final quark with mass
$m$ near the light cone in Breit frame, and $\Lambda^{2}$ is a
parameter in the IR cut-off. In the above two cases we set
$\lambda^{2}\sim\Lambda/Q$ and $\lambda\sim M/Q$, respectively.

First, we consider the on-shell case (a), according to the step
(i), the current in SCET$_{I}$ at leading order of $\lambda$ is
\cite{lb}
\begin{equation}j^{\mu}=[\bar{\xi}_{\bar{n}}W_{\bar{n}}]\gamma^{\mu}
\mathcal{C}(\mathcal{P}^\dagger,\bar\mathcal{{P}},\mu^{2})[W_{n}^{\dagger}\xi_{n}],
\end{equation}
where the reparameterization invariance(RPI) \cite{rpi} implies
Wilson coefficient
$$\mathcal{C}(\mathcal{P}^\dagger,\bar\mathcal{{P}},\mu^{2})
=\mathcal{C}(\mathcal{P}^\dagger\cdot\bar\mathcal{{P}},\mu^{2}).$$
Because EFT Lagrangian only takes coupling vertexes with the
fields involved interacting in a local way, there is no direct
coupling of collinear particles moving in the two separate
directions defined by $n$ and $\bar{n}$. However usoft modes can
still mediate between them. The calculations at one-loop level
give
\begin{equation}
\mathcal{C}(Q^{2},Q^{2})
=1+\frac{\alpha_{s}C_{F}}{4\pi}(-8+\frac{\pi^{2}}{6})
\end{equation}
and the UV renormalization factor for the current in SCET is
\begin{equation}
Z_{V}=1+\frac{\alpha_{s}C_{F}}{4\pi}
[\frac{2}{\epsilon^2}+\frac{3}{\epsilon}-\frac{2}{\epsilon}
\log(\frac{Q^2}{\mu^2})].
\end{equation}
Thus the RGE of Wilson coefficient $\mathcal{C}$ is
\begin{eqnarray}
\frac{d\log \mathcal{C}(Q^{2},\mu^2)}{d\mathrm{log}(\mu)}
=\gamma_{1}(\mu),
\end{eqnarray}
\begin{eqnarray}
\gamma_{1}(\mu) &\equiv& \mathbf{A}_{q}(\alpha_{s})\log
(\frac{Q^{2}}{\mu^{2}})+\mathbf{B}_{q}(\alpha_{s}) \nonumber\\
&=& -\frac{\alpha_{s}C_{F}}{4\pi} [4\log(\frac{\mu^2}{Q^2})+6].
\end{eqnarray}
In this paper we use the notions $\mathbf{A}$, $\mathbf{B}$ and
$\mathbf{C}$ in order to distinguish the well known coefficients
A, B and C in pQCD, but later we will find they are connected with
each other. So we have $\mathbf{A}_{q}^{(1)}=C_F $ and
$\mathbf{B}_{q}^{(1)}=-{3C_F}/{2}$ if $\mathbf{A}\equiv
\sum_{n}({\alpha_{s}}/{\pi})^{n}\mathbf{A}^{(n)},$ etc. With (6),
we can resumme terms such as double logarithmic from scale $\sim
\mathcal{O}(1)$ down to scale $\sim \mathcal{O}(\lambda)$, we
abbreviate this matching step as a chain
$QCD|_{Q^{2}}\longrightarrow SCET_I|_{Q^{2}\lambda^{2}}$.

According to the step (ii), usoft and collinear modes are
decoupled after field redefinitions. As a result,
\begin{equation}
\langle k|[\bar{\xi}_{\bar{n}}W_{\bar{n}}]\gamma^{\mu}
[W_{n}^{\dagger}\xi_{n}]|p\rangle\longrightarrow\langle
k|[\bar{\xi}_{\bar{n}}W_{\bar{n}}]|0\rangle\gamma^{\mu}
\langle0|T[Y_{\bar{n}}Y_{n}]|0\rangle
\langle0|[W_{n}^{\dagger}\xi_{n}]|p\rangle,
\end{equation}
where $T$ denotes time-ordered product.

According to the step (iii), note that the IR cut-off scale
$\Lambda^{2}>>p^{2}=k^{2}=m^{2}$, the transformed operator can
directly match onto the operator in SCET$_{II}$ associated with
re-scaling the external modes in SCET$_{I}$,
\begin{eqnarray}
\nonumber\ p_{c}=Q(0,1,0)\sim
Q(\lambda^{4},1,\lambda^2)\longrightarrow
p_{c}\sim Q(\eta^{2},1,\eta),\\
\nonumber\ p_{us}\sim
Q(\lambda^{2},\lambda^{2},\lambda^{2})\longrightarrow p_{s}\sim
Q(\eta,\eta,\eta).
\end{eqnarray}
In this case, there is no intermediated scale which separates
$SCET_I$ and $SCET_{II}$, and then the Wilson coefficient for this
step is one and the anomalous dimension is the same as SCET$_{I}$.
Although the RGE of Wilson coefficient still is (6), it runs from
the scale $\sim \mathcal{O}(\lambda)$ to the scale $\sim
\mathcal{O}(\eta)$. We will abbreviate this step as
$SCET_I|_{Q^{2}\lambda^{2}}\Rightarrow
SCET_{II}|_{Q^{2}\lambda^{4}}$. Collecting all the results above,
we obtain the well known Sudakov form factor $S(Q,\Lambda)$,
leaving other coefficients omitted,
\begin{equation}
S(Q,\Lambda)=\exp(-\int_{\Lambda}^{Q} \gamma_{1}(\mu)d\log\mu).
\end{equation}

For the off-shell case (b), it can be taken as a sub-diagram of
the on-shell case, from kinematical considerations, of which the
external legs are amputated. Thus, the step (i) is unchanged, but
in the step (ii) $\langle0|T[Y_{\bar{n}}Y_{n}]|0 \rangle$ changes
into
\begin{equation}
-\int_{0}^{\infty}\int_{0}^{\infty}dsdt
e^{iQ\lambda^{2}(s+t)}\langle
0|T[Y_{\bar{n}}(0,\bar{n}s)Y_{n}^{\dagger}(0,-nt)]|0\rangle,
\end{equation}
where $1/(Q\lambda^{2})$ is the effective contour length as shown
in \cite{ofsf} and
$$Y_{\bar{n}}(0,\bar{n}s)\equiv\mathrm{P}
\exp(ig\int_{0}^{s}d\beta\bar{n}\cdot A_{us}(\bar{n}\beta)),$$
$$Y_{n}(0,-nt)\equiv\mathrm{P}\exp(-ig\int_{0}^{t}d\beta n\cdot A_{us}(-n\beta)).$$
And in the step (iii) the jets with off-shellness
$-p^{2}=-k^{2}=M^2>>Q^2\lambda^4$ are integrated out in
SCET$_{II}$ and only Eq.(10) is left in the step (iii). Actually,
the running behaviors of Eq.(10) and $F_{IR}$ in \cite{onfsf} are
identical. The above steps adopted for (b) can be abbreviated as
$QCD|_{Q^2}\longrightarrow
SCET_{I}|_{Q^{2}\lambda^{2}}\longrightarrow
SCET_{II}|_{Q^{2}\lambda^{4}}$. Finally, Sudakov factor in the
off-shell case is
\begin{equation}
S(Q,Q\lambda^{2})=\exp(-\int_{M}^{Q}\gamma_{1}(\mu)d\log\mu+
\int_{Q\lambda^{2}}^{M}\gamma_{2}(\mu)d\log\mu),
\end{equation}
where $\gamma_{2}(\mu)=2\Gamma_{0}(g)-\log
({\mu^{2}}/({Q^{2}\lambda^{4}}))\Gamma_{\mathrm{cusp}}(g)=-\log
({\mu^{2}}/({Q^{2}\lambda^{4}}))\alpha_{s}C_{F}/\pi+\mathcal{O}(\alpha_{s}^2)$
as shown in \cite{onfsf}. Thus, SCET automatically derives the
results based on the analysis of separating momentum space into
collinear, soft and IR regions.

\section{$Q_T$ resummation of Higgs-boson production \cite{ktn}}

Now we focus on resummed parts of full transverse momentum
distribution of Higgs-boson produced via gluon fusion, while the
remaining terms corresponding to $Y$  \cite{ktt} and the
procedures of incorporating non-perturbative region ($Q_T \sim
\Lambda_{QCD}$) are omitted.

The dominant process for Higgs-bosons production at the Large
Hadron Collider(LHC) in the Standard Model are gluon fusion
through a heavy quark loop, mainly the top quark:
$p_1(P_{1})+p_2(P_{2})\rightarrow gg\rightarrow H(Q)+X$, with
$P_{1}=(0,2p,0)$ and $P_{2}=(2p,0,0)$. It is convenient to start
from the effective Lagrangian for Higgs-bosons and gluon couplings
\cite{h}:
\begin{equation}
\mathcal{L}_{Hgg} =h(\alpha_s(Q))G_{\mu\nu}^{a}G_{a}^{\mu\nu}H,
\end{equation}
where
$$h(\alpha_{s})=\frac{\alpha_{s}}{12\pi}(\sqrt{2}G_F)^{1/2}
[1+\frac{11}{4}\frac{\alpha_{s}}{\pi}].$$ From (12), the operator
for one Higgs-boson production is
$\mathcal{H}=G_{\mu\nu}^{a}G_{a}^{\mu\nu}$.

Taking into account Landau equation and IR power counting
\cite{ir}, the singular terms of $Q_T$ distribution in the limit
of $Q_T\rightarrow 0$ originate from soft and collinear modes,
which are emitted by partons from hadrons $p_1,p_2$. Thus, only
real gluon emission and the gluon-quark scattering diagrams
contribute to the singular parts of $Q_T$ distribution. In the
case of gluon emission, the emitted gluon can be soft or
collinear, while the quark can only be collinear. SCET is
appropriate to this semi-inclusive process with
$\lambda^{2}\sim{Q_T}/{Q},Q>>Q_T>>\Lambda_{QCD}$ and $Q\sim
m_{H}$. So the operator $\mathcal{H}$ can matches at the lowest
order of $\lambda$ onto
\begin{equation}
\mathcal{H}=\mathcal{G}_{\bar{n}}^{a\mu\nu}
\mathcal{C}(\mathcal{P}^\dagger\cdot\bar\mathcal{P},\mu^{2})
\mathcal{G}_{n\mu\nu}^a=\mathcal{G}_{\bar{n}}^{a\mu\nu}
\mathcal{C}(Q^2,\mu^{2}) \mathcal{G}_{n\mu\nu}^a,
\end{equation}
where
$$\mathcal{G}_{n}^{\mu\nu}=-({i}/{g})
W_{n}^{\dagger}[i\mathcal{D}_{n}^{\mu}+gA_{n}^{\mu},
i\mathcal{D}_{n}^{\nu}+gA_{n}^{\nu}]W_{n}.$$ And $n\leftrightarrow
\bar{n}$ for $\mathcal{G}_{\bar{n}}^{\mu\nu}$. The calculations at
one-loop level give
\begin{equation}
\mathcal{C}(Q^2,Q^2)=1+\frac{\alpha_{s}}{4\pi}(\mathcal{A}_{g}^{H}+\frac{C_A\pi^{2}}{2})
\end{equation}
and
\begin{equation}
Z_{\mathcal{H}}=1+\frac{\alpha_{s}}{4\pi}
[\frac{2C_{A}}{\epsilon^{2}}+\frac{2\beta_{0}}{\epsilon}
-\frac{2C_{A}}{\epsilon}\log(\frac{Q^{2}} {\mu^{2}})].
\end{equation}
Here, $C_A=N_c$, $\mathcal{A}_{g}^{H}=11+2\pi^{2}$,
$\beta_{0}=(11C_A-2n_f)/6$, and $n_f=5$ is the number of active
quark flavors. Thus, the RGE of $\mathcal{C}(Q^2,\mu^2)$ is
\begin{equation}
\frac{d\log\mathcal{C}(Q^2,\mu^{2})}{d\log(\mu)} =\gamma_{1}(\mu),
\end{equation}
where
\begin{eqnarray}
\gamma_{1}(\mu)&\equiv& \mathbf{A}_{g}(\alpha_{s})\log
(\frac{Q^{2}}{\mu^{2}})+\mathbf{B}_{g}(\alpha_{s}) \nonumber\\
&=& -(\frac{\alpha_{s}}{\pi})
[C_{A}\log(\frac{\mu^2}{Q^2})+\beta_{0}]
\end{eqnarray}
with $\mathbf{A}_{g}^{(1)}=C_A $ and
$\mathbf{B}_{g}^{(1)}=-\beta_{0}$. The evolution from the scale
$\sim \mathcal{O}(1)$ to the scale $\sim \mathcal{O}(\lambda)$
gives
\begin{equation}
\mathcal{C}(Q^2,Q^{2}\lambda^{2})=
\mathcal{C}(Q^2,Q^2)\exp(-\int_{Q^{2}\lambda^{2}}^{Q^{2}}
\frac{d\mu^{2}}{2\mu^{2}}\gamma_{1}(\mu)).
\end{equation}
Similar to the case of on-shell Sudakov form factor in last
section, Eq.(16) can be directly used to running $\mathcal{H}$
from $Q$ to $Q\lambda^{2}$ with no loss of degrees of freedom. For
example, the scaling of the gluon interacting with the incoming
quark is $\mathcal{O}(\eta^{2},1,\eta)$, which is constrained by
kinematics. After performing field redefinitions, the relevant
operator for Higgs-boson production at the scale $Q_T$, where
large logarithmic has been resummed by the RGE evolution in SCET,
is
\begin{eqnarray}
\mathcal{H}&=& \mathcal{C}(Q^2,Q^2\eta^{2})2\mathrm{Tr}
\{T[Y_{n}^{\dagger}Y_{\bar{n}}\mathcal{G}_{\bar{n}}^{\mu\nu}
Y_{\bar{n}}^{\dagger}Y_{n}\mathcal{G}_{n\mu\nu}]\}\nonumber \\
&=&
\mathcal{C}(Q^2,Q^2\eta^{2})T[\mathcal{Y}_{\bar{n}}^{ab}\mathcal{Y}_n^{ac}
\mathcal{G}_{\bar{n}}^{b\mu\nu}\mathcal{G}_{n\mu\nu}^c]\equiv
\mathcal{C}(Q^2,Q^2\eta^{2})\bar{\mathcal{H}},
\end{eqnarray}
where $\mathcal{Y}_{n(\bar{n})}$ is adjoint Wilson line from
$-\infty$ to $0$ in $n(\bar{n})$ direction for incoming fields.
Until now, we have completed the procedures corresponding to (i),
(ii) and (iii) mentioned in previous section. Next, with the
composite operator $\mathcal{H}$ at the renormalization scale
$\mu\sim Q\lambda^{2}$, we can relate $\mathcal{H}$ to
differential cross section at this scale:
\begin{equation}
\frac{1}{\sigma^{(0)}}\frac{{d\sigma}^{\mathrm{resum}}}
{dQ^{2}dydQ_{T}^{2}}=\frac{d}{dQ_{T}^{2}} \int_{0}^{Q_{T}^{2}}
dq_{T}^{2}e^{-\mathbf{S}_{g}(\mu,Q)}\mathcal{C}^{2}(Q^2,Q^2)
\frac{1}{\sigma^{(0)}}\frac{d\sigma^{\mathrm{SCET}}
(\mu)}{dQ^{2}dydq_{T}^{2}},
\end{equation}
where
\begin{eqnarray}
\sigma^{(0)}=(\sqrt{2}G_{F})\frac{\alpha_{s}^{2}(Q)m_{H}^{2}}
{576 S}\delta(Q^{2}-m_{H}^{2}),\qquad\ S=P_{1}^{-}P_{2}^{+},\\
\mathbf{S}_{g}(\mu,Q)=\int_{\mu^{2}}^{Q^{2}}
\frac{d\mu^{2}}{\mu^{2}}
[\mathbf{A}_{g}(\alpha_{s})\mathrm{log}(\frac{Q^{2}}{\mu^{2}})+\mathbf{B}_{g}(\alpha_{s})]
\end{eqnarray}
and ${d\sigma^{\mathrm{SCET}}(\mu)}/({dQ^{2}dydq_{T}^{2}})$
represents differential cross section calculated in $SCET_{II}$
with the composite operator $\bar{\mathcal{H}}$ at the
renormalization scale $\mu\sim Q\lambda^{2}$.

To proceed further, we use the following two facts: the soft and
collinear modes are decoupled in lagrangian at LO of $\lambda$ in
SCET$_{II}$; since the spins of hadron are summed over and a
colored octet operator vanishes between the color singlet states
\cite{hsf}, we have $\langle
p_{n}|\mathcal{B}_{n,\omega_1}^{a,\mu}
\mathcal{B}_{n,\omega_2}^{b,\nu}|p_{n}\rangle\propto
\delta^{ab}g_{\perp}^{\mu\nu} \langle
p_{n}|\mathrm{Tr}[\mathcal{B}_{n,\omega_1}^{\alpha}
\mathcal{B}^{n,\omega_2}_{\alpha}]|p_{n}\rangle$ with
$g_{\perp}^{\mu\nu}=g^{\mu\nu}-(n^{\mu}
\bar{n}^{\nu}+\bar{n}^{\mu} n^{\nu})/2$ and
$\mathcal{B}_{n\mu}^{a}=\bar{n}^{\nu}\mathcal{G}_{n\mu\nu}^{a}$.
Thus, the SCET cross section can be written as a factorized form:
\begin{equation}
\int_{0}^{Q_{T}^{2}} dq_{T}^{2}\mathcal{C}^{2}(Q^2,Q^2)
\frac{1}{\sigma^{(0)}}\frac{d\sigma^{\mathrm{SCET}}(\mu)}{dQ^{2}dydq_{T}^{2}}
=g_{p_1}(x_{1},\mu)g_{p_2}(x_{2},\mu),
\end{equation}
where $x_{1}={Qe^{y}}/ \sqrt{S}, x_{2}={Qe^{-y}}/\sqrt{S}$ for
$Q_{T}^{2}<<Q^{2}$. Because of KLN theorem, the contributions from
the soft modes are free of IR divergences. So only the collinear
divergences are survived. Through matching the SCET cross section
onto a product of two PDFs given by \cite{hsf}, which are
equivalent to the conventional PDFs
$f_{a/p_{1(2)}}(\xi_{1(2)},\mu)$ at leading order of $\lambda$,
the remained IR divergences can be absorbed into these
nonperturbative inputs \cite{dy}, of which the evolutions are
controlled by the DGLAP equations. This step of matching Eq.(23)
onto $\sum_{a,b}f_{a/p_{1}}(\xi_{1},\mu)f_{b/p_{2}}(\xi_{2},\mu)$
leads to
\begin{eqnarray}
g_{p_{1(2)}}(x_{1(2)},\mu)&\equiv&
\sum_{a=g,q}(f_{a/p_{1(2)}}\otimes
\mathbf{C}_{ga})(x_{1(2)},\mu) \nonumber\\
&=&\sum_{a}\int_{x_{1(2)}}^{1}\frac{d\xi}{\xi}
f_{a/p_{1(2)}}(\xi,\mu) \mathbf{C}_{ga}(\frac{x_{1(2)}}{\xi},\mu),
\end{eqnarray}
Obviously, at the tree-level only
$\mathbf{C}_{gg}^{(0)}(z)=\delta(1-z)\neq 0$. To extract
$\mathcal{O}(\alpha_s)$ coefficients $\mathbf{C}_{ga}^{(1)}$, we
can take a short-cut. As demonstrated in \cite{sr}, EFT is the
language of method of regions, and we can use the results of
\cite{pt}, which are obtained by the method of regions in pQCD
framework, to simplify our calculations. The only difference
between the pQCD and SCET, if we set the factorization scale
$\mu_{F}=\mu$ in both theories, is that the choices of
renormalization scale in pQCD \cite{pt} is $\mu=Q$, while in SCET
is $\mu\sim Q\eta$. Fortunately, this discrepancy is compensated
by the expansion of Sudakov factor $\exp(-\mathbf{S}_{g}(\mu,Q))$,
and the results for $\mathcal{O}(\alpha_s)$ coefficients
$\mathbf{C}_{ab}^{(1)}$ are the same as the known formula
\cite{pt} in pQCD approach, i.e.,
\begin{eqnarray}
\mathbf{C}_{ga}^{(1)}(z)&=&-\frac{1}{2}P_{ga}^{\epsilon}(z)+\frac{1}{4}\delta_{ga}
\delta(1-z)(C_{A}\frac{\pi^{2}}{3}+\mathcal{A}_{g}^{H}),
\end{eqnarray}
where the scale $\mu$ has been set to $Q_T$ in order to minimize
the logarithmic in $\mathbf{C}_{ga}^{(1)}(z)$, and the
approximation $\alpha_s(Q)=\alpha_s(Q_T)$ is taken up to NNLLO in
$\mathcal{C}^{2}(Q^2,Q^2)$. And $P_{ga}^{\epsilon}(z)$ represent
the $\mathcal{O}(\epsilon)$ terms in the DGLAP splitting kernels,
which are given by
\begin{eqnarray}
\nonumber P_{gq}^{\epsilon}(z)&=&-C_{F}z, \nonumber\\
P_{gg}^{\epsilon}(z)&=&0. \nonumber
\end{eqnarray}

Combining Eq.(21)-(24), we obtain resummation formula for
transverse momentum distributions of Higgs-bosons production in
SCET:
\begin{equation}
\frac{1}{\sigma^{(0)}}\frac{{d\sigma}^{\mathrm{resum}}}
{dQ^{2}dydQ_{T}^{2}}=\frac{d}{dQ_{T}^{2}} \sum_{ab}
e^{-\mathbf{S}_{g}(Q_{T},Q)}(f_{a/p_{1}}\otimes
\mathbf{C}_{ga})(x_{1},Q_{T})(f_{b/p_{2}}\otimes
\mathbf{C}_{gb})(x_{2},Q_{T}).
\end{equation}
This form is different from the known CSS formula in $\vec{b}$
space \cite{kt,ktt,abc}:
\begin{eqnarray}
\frac{1}{\sigma^{(0)}}\frac{{d\sigma}^{\mathrm{resum}}}
{dQ^{2}dydQ_{T}^{2}}=\int_{0}^{\infty} \frac{db}{2\pi}
bJ_0(bQ_T)\sum_{ab} e^{-S_{g}(\frac{c}{b},Q)}(f_{a/p_{1}}\otimes
C_{ga})(x_{1},\frac{c}{b})(f_{b/p_{2}}\otimes
C_{gb})(x_{2},\frac{c}{b}),\nonumber\\
\end{eqnarray}
where $c=2e^{-\gamma_{E}}$ and $\gamma_{E}$ is Euler's constant.
As we know, the reason of using the $\vec{b}$ space in pQCD is to
preserve conservation of momenta, while in SCET the conservation
of momenta is automatically satisfied because the matrix elements
is calculated at scale $Q_T$ with the matched effective operator.
Previously, a similar formula has been derived by the authors of
\cite{ddt}, which is incomplete, only at NLLO, and corresponds to
our result of setting $\mathbf{C}_{ga}=\delta_{ga}\delta(1-z)$ in
Eq.(26). Later, the authors in \cite{ddt2} generalized the DDT
formula to NNLLO with the CSS formula, in which its coefficients
$\tilde{A}$, $\tilde{B}$ and $C$  at $\mathcal{O}(\alpha_{s})$ are
just our $\mathbf{A}^{(1)}$, $\mathbf{B}^{(1)}$ and
$\mathbf{C}^{(1)}$. We also expect that they would agree at higher
orders, and thus we propose that at NNLLO
$\mathbf{A}^{(2)}=\tilde{A}^{(2)}=A^{(2)}$ and
$\mathbf{B}^{(2)}=\tilde{B}^{(2)}=B^{(2)}+2(A^{(1)})^2\zeta(3)$,
if all the coefficients are defined by $A=\sum _n
(\frac{\alpha_s}{\pi})^{n}A^{(n)}$ etc. The first relation has
been noticed by the author of \cite{dis}. The connections between
$\mathbf{A}$, $\mathbf{B}$ and $\mathbf{C}$ in SCET and A, B and C
in pQCD, therefore, are established through this way. The
numerical results of the $Q_T$ resummation formula in SCET and
that of traditional CSS formalism in pQCD are in agreement at
NNLLO, for details we suggest to refer to \cite{ddt2}.

We summarize the steps for $Q_T$ resummation with a chain:
$$QCD|_{Q^{2}}\longrightarrow SCET_{I}|_{Q^{2}\lambda^{2}}\Rightarrow
SCET_{II}|_{Q^{2}\eta^{2}>>\mu_{0}^{2}}\longrightarrow
DGLAP|_{\mu_{0}^{2}},$$ where the last arrow indicates that the
evolutions below $SCET_{II}$ are governed by the DGLAP equations.

\section{discussion}

$\bullet$ Using the above methods to the production of lepton pair
via virtual photon, we can get the similar result as Eq.(26)
corresponding to that of \cite{abc}.

$\bullet$ From our above analysis, it can be seen that SCET
provides a natural framework of $Q_T$ resummation by conventional
RGE in EFT.

$\bullet$ Because of the differential form of Eq.(26) in $Q_T$
space, it provides a simple and natural approach to numerical
calculation and covering the effects in non-perturbative region
with $Q_T\sim \Lambda_{QCD}$ \cite{ddt2}.

$\bullet$ The reformulation of joint resummation can also be made
straightforwardly in SCET. In fact, the conclusions of threshold
resummation for ${d\sigma^{\mathrm{resum}}}/{dQ^2}$ in moments
$\bar{N}=e^{\gamma_{E}}N$ space for Higgs-boson production
processes can be shown by the chain\cite{dy}: in the case of
$z\rightarrow1$ and $\lambda^{2}=1-z \sim \frac{1}{\bar{N}}$,
$$QCD|_{Q^{2}}\longrightarrow SCET_{I}|_{Q^{2}\lambda^{2}}\Rightarrow
SCET_{II}|_{Q^{2}\eta^{2}>>\mu_{0}^{2}}\longrightarrow
DGLAP|_{\mu_{0}^{2}}.$$ We observe that the two chains of
Higgs-boson production processes in SCET have identical structure.
This suggests that we can do threshold and $Q_T$ resummation for
${d\sigma^{\mathrm{resum}}}/{dQ^2dQ_T^2}$ simultaneously. The
relevant scale $\lambda^{2}=1/\chi(\bar{N},\bar{b}\equiv
bQe^{\gamma_{E}}/2)$ of \cite{jr} is replaced by the
$1/\chi(\bar{N},Q/Q_T)$ in SCET, which is an interpolation between
$\lambda^{2}\sim \frac{1}{\bar{N}}$ and $\lambda^{2}\sim Q_T/Q$.
That is to say
$$\chi(\bar{N},Q/Q_T)=Q/Q_T+\frac{\bar{N}}{1+1/(4\bar{N})Q/Q_T}$$
approaches to $\bar{N}$ for $Q/Q_T<<\bar{N}$ and to $Q/Q_T$ for
$Q/Q_T>>\bar{N}$, respectively. The matching steps for joint
resummation then can be written as
$$QCD|_{Q^{2}}\longrightarrow SCET_{I}|_{Q^{2}/\chi}\Rightarrow
SCET_{II}|_{Q^{2}/\chi^{2} >>\mu_{0}^{2}}\longrightarrow
DGLAP|_{\mu_{0}^{2}},$$ which leads to similar result as Eq.(26)
corresponding to that of \cite{jr}.

\section{conclusion}

We have shown $Q_T$ resummation of Higgs-boson production in the
framework of SCET and given a simple relationship between
anomalous dimension of operator in SCET and
$\mathcal{O}(\alpha_s)$ coefficients A and B in pQCD, and
$\mathcal{O}(\alpha_s)$ coefficient C is reproduced by matching
the matrix elements of hadrons at the scale $Q_T$ onto a product
of two PDFs in $SCET_{II}$, and their connections at higher order
are obtained by comparing the formulas in the two frames. We also
show that the reformulation of joint resummation can be performed
in SCET straightforwardly. Finally, we want to point out that any
processes, which are limited to the soft and collinear regions by
kinematics, can be treated in SCET following the steps outlined
above, and moreover, for the high energy hard scattering processes
involving the colored final state which is neither soft in the
sense of SCET-including the heavy quark effective theory (HQET)
nor collinear, the method for
$Q_T$ resummation based on SCET collapses.\\
\textbf{Acknowledgments}: This work is supported in part by the
National Natural Science Foundation of China and the Specialized
Research Fund for the Doctoral Program of Higher Education.


\begin{thebibliography}{xx}

\bibitem{bsr}
C.W.Bauer, S. Fleming and M.Luke,Phys.Rev.\textbf{D63},
014006(2001).
\bibitem{lb}
C.W.Bauer, D. Pirjol and I.W.Stewart, Phys.Rev.\textbf{D65},
054022(2002).
\bibitem{hsf}
C.W.Bauer, S.Fleming, D.Pirjol, I.Z.Rothstein and I.W.Stewart,\\
Phys.Rev.\textbf{D66}, 014017(2002).
\bibitem{rpi}
C.W.Bauer, D. Pirjol and I.W.Stewart, Phys.Rev.\textbf{D68},
034021(2003).
\bibitem{ee}
C.W.Bauer, A.V.Manohar and M.B.Wise, Phys.Rev.Lett.\textbf{91},
122001(2003).
\bibitem{eet}
C.W.Bauer, C.Lee, A.V.Manohar and M.B.Wise, Phys.Rev.\textbf{D70},
034014(2004).
\bibitem{dis}
A.V.Manohar, Phys.Rev.\textbf{D68}, 114019(2003).
\bibitem{dy}
A.Idilbi and X.D.Ji, hep-ph/0501006.
\bibitem{ktn}
Yang Gao, Chong Sheng Li and Jian Jun Liu, hep/ph-0501229.
\bibitem{kt}
R.P.Kauffman, Phys.Rev.\textbf{D44}, 1415(1991).
\bibitem{ktt}
E.L.Berger and J.W.Qiu, Phys.Rev.\textbf{D67}, 034026(2003).
\bibitem{sr}
M.Beneke and V.A.Smirnov, Nucl.Phys.\textbf{B522}, 321(1998);\\
M.Beneke, A.P.Chapovshy, M.Diehl and T.Feldmann,
Nucl.Phys.B\textbf{643}, 431(2002).
\bibitem{pt}
D.de Florian, M.Grazzini, Phys.Rev.Lett.\textbf{85}, 4678(2000);
Nucl.Phys.\textbf{B616}, 247(2001).
\bibitem{onfsf}
G.P.Korchemsky, Phys.Lett.\textbf{B219}, 330(1989);
\emph{ibid.}\textbf{B220}, 629(1989).
\bibitem{ofsf}
G.P.Korchemsky and A.V.Radyushkin, Phys.Lett.\textbf{B279},
359(1992).
\bibitem{h}
M.Spira, A.Djouadi, D.Graudenz, P.M.Zerwas,
Nucl.Phys.\textbf{B281}, 310(1987).
\bibitem{ir}
G.Sterman, Phys.Rev.\textbf{D17}, 2773(1978).
\bibitem{abco}
J.C.Collins and D.E.Soper, Nucl.Phys.\textbf{B193}, 381(1981);
\emph{ibid.}\textbf{B197}, 446(1982).
\bibitem{abc}
J.C.Collins, D.E.Soper and G.Sterman, Nucl.Phys.\textbf{B261},
104(1985).
\bibitem{ddt}
Yu.L.Dokshizer, D.I.D'Yakanov and S.I.Troyan,
Phys.Lett.\textbf{B78}, 290(1978).
\bibitem{ddt2}
R.K.Ellis and S.Veseli, Nucl.Phys.\textbf{B511},649(1998).
\bibitem{jr}
A.Kulesza, G.Sterman and W.Vogelang, Phys.Rev.\textbf{D69},
014012(2004);\\
\emph{ibid.} \textbf{D66}, 014011(2002).
\end{thebibliography}
\end{document}